\documentclass[epj]{svjour}
\usepackage[fleqn]{amsmath}
\usepackage{amssymb} % only for \lesssim, \gtrsim
\usepackage{graphicx}

\def\H{\mathcal H}
\def\abs#1{\lvert#1\rvert}
\def\no#1{:#1:}
\DeclareMathOperator{\erf}{erf}

\begin{document}
\title{Superfluidity near Phase Separation in Bose-Fermi Mixtures}
\author{Tilman Enss \and Wilhelm Zwerger}
\institute{Physik Department, Technische Universit\"at M\"unchen,
  D-85747 Garching, Germany}
\date{Received: date / Revised version: date}
\abstract{%
  We study the transition to fermion pair superfluidity in a mixture
  of interacting bosonic and fermionic atoms.  The fermion interaction
  induced by the bosons and the dynamical screening of the condensate
  phonons due to fermions are included using the nonperturbative
  Hamiltonian flow equations.  We determine the bosonic spectrum near
  the transition towards phase separation and find that the superfluid
  transition temperature may be increased substantially due to phonon
  damping.
}
\maketitle
%
%%%%%%%%%%%%%%%%%%%%%%%%%%%%%%%%%%%%%%%%%%%%%%%%%%%%%%%%%%%%%%%%%%%%%%%%
\section{Introduction}
\label{intro}
%%%%%%%%%%%%%%%%%%%%%%%%%%%%%%%%%%%%%%%%%%%%%%%%%%%%%%%%%%%%%%%%%%%%%%%%
Mixtures of bosonic and fermionic atoms have initially been used for
sympathetic cooling of fermions.  This allows reaching the degenerate
regime in ultracold Fermi gases despite the freezing out of
thermalizing collisions between fermions in a single internal state at
low temperature \cite{group1,taglieber2008qdt}.  From a theory point
of view, Bose-Fermi mixtures are of interest in their own right.
Indeed, already at the mean-field level, a number of different ground
states have been predicted.  Depending on the density $n_F$ of
fermions and the value of the Bose-Fermi ($a_{BF}$) and Bose-Bose
($a_{BB}$) scattering lengths, a phase separation instability is
expected for strong repulsive Bose-Fermi interactions $a_{BF}^2\gtrsim
a_{BB}/n_F^{1/3}$ \cite{group2}.  For sufficiently strong attractive
interactions, in turn, the mixture is unstable.  Accordingly, in a
trap, a collapse is expected beyond a critical value of the particle
number \cite{group3}.  The situation $a_{BF}<0$ applies, for instance,
to the case of $^{40}$K-$^{87}$Rb mixtures, where signatures for a
collapse have been observed experimentally
\cite{modugno2002cdf,ospelkaus2006idd,endnote1}.  The scattering
length $a_{BF}$ can be controlled using magnetically tunable Feshbach
resonances \cite{group4}.  This opens the possibility to explore a
number of novel phases in Bose-Fermi mixtures with nontrivial
many-body correlations. One of the simplest among those is an $s$-wave
superfluid in a Fermi gas with two internal states, where the
attractive interaction is mediated by the exchange of phonons in the
condensed bosonic gas \cite{heiselberg2000iii,group4a}.  More exotic
states like $p$-wave pairing \cite{group5} or an odd-frequency
$s$-wave state \cite{kalas2008ofp} may arise in the case where only a
single internal state of the fermionic atoms is present.  For
Bose-Fermi mixtures on a lattice, a number of nontrivial phases have
been predicted, for instance a supersolid phase of bosons in the
presence of a fermionic density wave \cite{group6} or phases with
composite fermions \cite{lewenstein2004abf}.

In our present work, we reconsider the rather basic problem of induced
pairing in a balanced gas of fermions with two internal states.  The
bosons are assumed to form a Bose-Einstein condensate (BEC) whose
density fluctuations can be described by the standard Bogoliubov
theory. Even in the absence of a direct interaction between the
fermionic atoms there is an induced interaction mediated by the
phonons of the BEC.  This situation is analogous to that of
phonon-mediated superconductivity of electrons in a solid.  As derived
in textbooks \cite{ashcroft1976ssp}, second order perturbation theory
in the Bose-Fermi coupling leads to an attractive, retarded
interaction between the fermions in the $s$-wave channel
\cite{frohlich1952iel}.  In weak coupling, the resulting BCS
instability appears at a temperature $T_c$ much smaller than the Fermi
temperature $T_F$.  The effects of retardation are then negligible
because pairing only affects fermions at the Fermi surface
\cite{endnote2}.  In the context of cold gases, however, reaching
temperatures far below $T_F$ is impossible.  The observation of
fermionic superfluidity requires $T_c/T_F$ to be of order $0.1$ or
larger, similar to the situation of attractive Fermi gases near a
Feshbach resonance
\cite{ketterle2008mpa,bloch2008mbp,giorgini2008tuf}.  To study induced
pairing in Bose-Fermi mixtures, weak coupling approaches in which only
the physics in the vicinity of the Fermi surface is relevant, are
therefore not reliable.  This applies in particular in the regime
close to the instability for phase separation where mean-field
calculations predict critical temperatures that are favorable for
observation \cite{heiselberg2000iii}.  Indeed, the strong interactions
and the fact that---in contrast to the situation in solids---bosons
and fermions have comparable masses, lead to a strong renormalization
of the phonon modes (dynamical screening) which is due to the
excitation of fermionic particle-hole pairs.  This effect lowers the
phonon frequencies and also gives rise to damping, thus broadening the
phonon spectral function.  If the dimensionless Bose-Fermi coupling
exceeds a critical value, the phonon frequencies become negative and
the velocity of sound imaginary.  This signals an instability towards
phase separation (for repulsive Bose-Fermi interaction) or collapse
(for attractive interaction). Since softer phonons induce a stronger
fermion interaction, we concentrate on the parameter region near this
instability, which is the most favorable for fermion pair
superfluidity.  Obviously, the presence of strong fluctuations which
may increase $T_c$ requires a technique that goes beyond the
mean-field level of a BCS approach, where retardation and the
backaction between the fermions and the phonons of the reservoir (in
our case the BEC) are negligible.

In conventional superconductors, the effects of retardation are
usually treated within Eliashberg theory \cite{abrikosov1963mqf}.
This approach typically relies on the assumption that the induced
interaction between fermions still appears only close to the Fermi
surface and also that the Fermi velocity is much larger than the
velocity of sound (adiabatic limit). In the context of cold atoms,
however, where $T_c/T_F$ may be of order unity, fluctuations with
energies up to the Fermi energy $E_F$ become important.  Moreover, in
typical mixtures like $^{40}$K and $^{87}$Rb, the mass ratio of
fermions and bosons is of order unity.  Provided that the healing
length of the BEC is of the same order as the average spacing between
fermions, the resulting ratio of the sound and Fermi velocities is
near unity.  A treatment of induced fermion pairing in Bose-Fermi
mixtures in terms of Eliashberg theory in such a situation was given
by Wang \cite{wang2006sct}.  He found that the strong-coupling effects
enhance $T_c$ considerably, at least in the regime $\abs{a_{BF}}k_F
\lesssim 0.1$.

To deal with the problem of strong coupling in Bose-Fermi mixtures, we
propose a different route using a renormalization group method which
automatically takes into account effects on all energy scales and
naturally identifies the elementary excitations of the coupled system.
To this end, we follow Wegner's idea of a continuous unitary
transformation of the Hamiltonian \cite{wegner1994feh,kehrein2006fea}.
It may be viewed as a reorganization of perturbation theory in such a
way that the new basis describes dressed particles which do not decay
and whose effective interactions are regular.  In this way, the
induced interaction is always attractive, free of singularities and
truly retarded (i.e., it vanishes for large energy transfer or short
times) \cite{lenz1996fee,mielke1997cct}. The method thus properly
describes the effective interactions even in a regime where the
typical fermion energies exceed the phonon energy.  By retaining the
full Hilbert space it allows to deal with fluctuations far away from
the Fermi surface in a natural manner and provides quantitatively
reliable results for the critical temperature near the interesting
regime of the phase separation instability.  Compared with Eliashberg
theory, the merit of the flow equations is that they yield a
block-diagonal Hamiltonian from which one can read off the elementary
excitations and identify the relevant degrees of freedom.  Moreover,
in contrast to the Wilson momentum-shell RG where scattering between
bare particles depends on both frequencies and momenta, the scattering
of renormalized particles in the flow equations depends on momenta
alone, which makes them easier to parametrize numerically.

The plan of the paper is as follows: in section \ref{sec:method} we
introduce the Hamiltonian flow equation method which allows us to
separate energy scales and avoid spurious singularities.  In section
\ref{sec:results} we present our results for the renormalization of
the phonon dispersion relation near the transition towards phase
separation (where the effect is largest), and for the change in the
induced interaction and of $T_c$ due to this damping.  Finally, we
discuss our results and where best to observe pair superfluidity
experimentally in section \ref{sec:discuss}.

%%%%%%%%%%%%%%%%%%%%%%%%%%%%%%%%%%%%%%%%%%%%%%%%%%%%%%%%%%%%%%%%%%%%%%%%
\section{Hamiltonian flow equations}
\label{sec:method}
%%%%%%%%%%%%%%%%%%%%%%%%%%%%%%%%%%%%%%%%%%%%%%%%%%%%%%%%%%%%%%%%%%%%%%%%
We consider a homogeneous, three-dimensional Bose-Fermi mixture of
spin-polarized bosons and fermions in two equal\-ly populated hyperfine
states.  The bosons interact via a repulsive pseudopotential with
strength $g_{BB}=4\pi\hbar^2 a_{BB}/m_B$, where $a_{BB} > 0$ is the
associated $s$-wave scattering length and $m_B$ the boson mass.  The
interaction $g_{BF}=2\pi \hbar^2 a_{BF}/m_r$ between bosons and
fermions with reduced mass $m_r$ may be attractive or repulsive,
depending on the sign of the interspecies scattering length $a_{BF}$.
For simplicity, we assume that the direct interaction between fermions
is negligible.

At low temperatures the bosons are condensed and it is sufficient to
consider how the fermions couple to the phononic excitations on top of
the condensate. The resulting model Hamiltonian (more precisely
$\H-\mu N$)
\begin{align}
 \label{eq:fro}
 \H & = \H_0 + \H_{\text{int}} \\
 \H_0 & = \sum_q \omega_q \no{a_q^\dagger a_q}
 + \sum_k \epsilon_k \no{c_k^\dagger c_k} \notag \\
 H_{\text{int}} & = \sum_{kq} M_q \left( a_{-q}^\dagger + a_q \right)
 c_{k+q}^\dagger c_k \notag
\end{align}
then coincides with that of an idealized description of the
electron-phonon interaction in solids.  Here $k=\{\vec k,\sigma\}$
denotes the momentum and effective spin degree of freedom
$\sigma=\uparrow,\downarrow$ to label the different hyperfine states.
Similarly, $a^{(\dagger)}$ and $c^{(\dagger)}$ are creation and
annihilation operators for the phonons and fermions, respectively,
while $\no{\ldots}$ denotes normal ordering.  The single-particle
energies of the fermions are $\epsilon_k = k^2/2m_F-\mu_F$.  The
phonons of the BEC are described by a Bogoliubov spectrum of the form
$\omega_q = c_s\abs q \sqrt{1+q^2\xi^2}$ with phonon velocity
$c_s=\sqrt{n_B g_{BB}/m_B}$ and healing length
$\xi=1/\sqrt{4m_Bn_Bg_{BB}}$ for the condensate with density $n_B$.
The fermion-phonon coupling is given by $M_q = g_{BF}
\sqrt{2n_B\omega_q^0/\omega_q}$ with bare bosonic dispersion
$\omega_q^0=q^2/2m_B$.

On the mean-field level the stability of a Bose-Fermi mixture is
guaranteed by the condition \cite{pethick2002bec}
\begin{align}
  \label{eq:lambdamf}
  \lambda_{\text{MF}} = \frac{\gamma_{BF}^2}{\gamma_{BB}} < 1
\end{align}
with the dimensionless couplings $\gamma_{BF}=g_{BF} N(E_F)$ and
$\gamma_{BB}=g_{BB} N(E_F)$, where $N(E_F)$ is the bare fermionic
density of states at the Fermi level.  Even for weak Bose-Fermi
coupling $\gamma_{BF}$ the system can become unstable if $\gamma_{BB}$
is also small (for typical values see section \ref{sec:results}
below).  As we will see below in equation \eqref{eq:Vfermi}, the
induced fermionic coupling $\lambda$ at weak-coupling is given by the
effective Cooper pair interaction at the Fermi surface averaged over
the scattering angle,
\begin{align}
  \label{eq:lambda}
  \lambda = \gamma_{BF}^2 x \log\Bigl(1+\frac{1}{x\gamma_{BB}}\Bigr)
\end{align}
with $x=n_Bm_B/(3n_Fm_F)$ relating the densities and masses of
bosonic and fermionic atoms.  Note that in the limit $x\gamma_{BB}\gg
1$ (small healing length $\xi\ll k_F^{-1}$) the angular average has no
effect and $\lambda \to \lambda_{\text{MF}}$.  We are interested in
the regime where $\lambda$ is of order unity: in this case the
polaronic effects described by an enhanced effective fermion mass
\cite{mahan2000mpp}
\begin{align}
  \frac{m^*}{m} = 1+\lambda
\end{align}
are not very large (we will consider them in a future publication
\cite{inprep}).  However, at the same time the induced fermionic
interaction $\lambda$ is quite large compared to the very weak bare
fermion repulsion (see section \ref{sec:results} below).

The standard Fr\"ohlich transformation
\cite{frohlich1952iel,abrikosov1963mqf} decouples the fermionic and
bosonic degrees of freedom and yields an induced fermion-fermion
interaction.  In particular, for the BCS channel $k'=-k$ one obtains
\begin{align}
  \label{eq:VFr}
  V_{k,-k,q}^{\text{(Fr)}} = -\frac{\omega_q M_q^2}
  {\omega_q^2 - (\epsilon_{k+q}-\epsilon_k)^2}
\end{align}
which has a singularity if large fermion energies are relevant in the
gap equation, as happens in the case of cold atoms.  A
well-established way to avoid such divergences is to perform a
regularization and renormalization that first takes into account
scattering processes with large energy transfer (off-shell) and
successively proceeds to processes with smaller energy transfer
(approaching on-shell scattering).  In this way, perturbation theory
is reorganized and resummed so as to satisfy energy scale separation
and to avoid small energy differences in the denominator of
perturbative expressions \cite{kehrein2006fea}.

Essentially, there is some freedom in choosing the unitary
transformation to decouple the fermion and phonon sectors: real
physical processes (on-shell) of course have to remain unchanged, but
the off-shell interaction (which is important for fermion pairing) can
be made regular by choosing an appropriate basis for the fermionic
quasiparticles.

Specifically, this change of basis is achieved by a continuous unitary
transformation on the Hamiltonian that eliminates scattering processes
with successively lower energy transfer
\cite{wegner1994feh,kehrein2006fea}.  It can be expressed in the form
of a differential flow equation
\begin{align}
  \label{eq:flow}
  \frac{d\H(\ell)}{d\ell} = [\eta(\ell),\H(\ell)]
\end{align}
with a flow parameter $\ell$ going from $0$ to $\infty$ that has
dimension $(\text{energy transfer})^{-2}$.  Here $\eta(\ell)$ is an
anti-hermitian operator which generates the unitary transformation 
\begin{align}
  U(\ell) = T_\ell \exp\left( \int_0^\ell d\ell' \, \eta(\ell') \right)
\end{align}
(with $\ell$-ordering defined in the same way as time ordering) such
that
\begin{align}
  \H(\ell) = U(\ell) \H(\ell=0) U(\ell)^\dagger \;.
\end{align}
There is some freedom in choosing $\eta$ appropriately; for models
where the Hamiltonian can be split into diagonal and interacting parts
Wegner \cite{wegner1994feh} suggested the canonical choice
\begin{align}
  \eta(\ell) = [\H_0(\ell),\H_{\text{int}}(\ell)]
\end{align}
which makes the Hamiltonian increasingly energy diagonal along the
flow and guarantees energy scale separation.  For the Fr\"ohlich
Hamiltonian \eqref{eq:fro} we choose the generator that has been used
originally by Lenz and Wegner \cite{lenz1996fee}
\begin{align}
  \eta(\ell) = \sum_{kq} \left( M_{kq} \alpha_{kq} a_{-q}^\dagger +
    M_{k+q,-q} \beta_{kq} a_q \right) c_{k+q}^\dagger c_k
\end{align}
where
\begin{align}
  \alpha_{kq} & = \epsilon_{k+q}-\epsilon_k+\omega_q & 
  \beta_{kq} & =  \epsilon_{k+q}-\epsilon_k-\omega_q 
\end{align}
denote the energy gain in the fermion-phonon scattering (on-shell
$\alpha_{kq}=\beta_{kq}=0$).  The flow equation \eqref{eq:flow} leads
to a flow of the single-particle energies $\epsilon_k$ and $\omega_q$
and the fermion-phonon coupling $M_{kq}$.  Moreover, it generates a
new coupling $V_{kk'q}$ between two fermions.  Two-phonon terms are
absent under the assumption of harmonic density waves
\cite{frohlich1952iel}, and higher couplings between several phonons
and fermions would only be generated at higher orders in the
fermion-phonon coupling beyond $\mathcal O(M^2)$.  Following
\cite{lenz1996fee} we choose to neglect these and consider only the
following running couplings in the Hamiltonian:
\begin{align}
  \H(\ell) & = \sum_q \omega_q(\ell) \no{a_q^\dagger a_q} \\
  & \quad + \sum_k
  \Bigl(\epsilon_k(\ell) - 2\sum_q n_{k+q} V_{k,k+q,q}(\ell)\Bigr)
  \no{c_k^\dagger c_k} \notag \\
  & \quad + \sum_{kk'q} V_{kk'q}(\ell)
  \no{c_{k+q}^\dagger c_{k'-q}^\dagger c_{k'} c_k} \notag \\
  & \quad + \sum_{kq} \left( M_{kq}(\ell) a_{-q}^\dagger +
    M_{k+q,-q}(\ell) a_q \right) c_{k+q}^\dagger c_k \notag \\
  & \quad + \text{irrelevant terms.} \notag 
\end{align}
The renormalization of these couplings up to second order $\mathcal
O(M^2)$ in the flowing fermion-phonon coupling is given by the flow
equations
\begin{align}
  \frac{dM_{kq}}{d\ell} & = -\alpha_{kq}^2 M_{kq} \\
  \frac{d\omega_q}{d\ell} & = 2 \sum_k M_{kq}^2 \alpha_{kq}
  \left(n_{k+q}-n_k\right) \\
  \frac{d\epsilon_k}{d\ell} & = -2\sum_q \bigl( n_q M_{k+q,-q}^2
    \beta_{kq} + (n_q+1) M_{kq}^2 \alpha_{kq} \bigr) \\
  \frac{dV_{kk'q}}{d\ell} & = M_{kq} M_{k'-q,q} \beta_{k',-q} -
  M_{k+q,-q} M_{k',-q} \alpha_{k',-q}
\end{align}
where all couplings on the right-hand side are $\ell$-dependent and
$n_k$ and $n_q$ are the fermionic and bosonic occupation numbers,
resp.  The initial conditions are $\epsilon_k(\ell=0) = \epsilon_k$,
$\omega_q(\ell=0) = \omega_q$, $M_{kq}(\ell=0)=M_q$ and
$V_{kk'q}(\ell=0)=0$ (without direct fermion-fermion interaction).
The flow equation for the fermion-phonon coupling can be solved
immediately,
\begin{align}
  M_{kq}(\ell) = M_q \exp\Bigl(-\int_0^\ell d\ell'
  \alpha_{kq}^2(\ell')\Bigr) \,.
\end{align}
$M_{kq}$ vanishes during the flow as the unitary transformation
successively decouples the fermion and boson sectors of the
Hamiltonian.  Since we assume a weak fermion-phonon coupling we
neglect the renormalization of the fermion single-particle energies
$\epsilon_k$.  Instead, we are mainly interested in the effect of
phonon damping on the induced interaction $V$, to see whether already
a small damping may change the interaction in the limit of resonant
scattering.  The flow of the phonon energies in three dimensions then
becomes
\begin{align}
  \frac{d\omega_q}{d\ell} & = -\frac{M_q^2}{2\pi^2} \int_0^\infty dk
  \, k^2n_k \int_{-1}^1 d(\cos\theta) \notag \\
  & \qquad \left( \alpha_{kq} e^{-2\int_0^\ell d\ell'
      \alpha_{kq}^2} + \beta_{kq} e^{-2\int_0^\ell d\ell'
      \beta_{kq}^2} \right) \,.
\end{align}
The integration over the angle $\theta$ between $\vec k$ and $\vec q$
is lengthy but straightforward,
\begin{align}
  \frac{d\omega_q}{d\ell} & = -\frac{M_q^2m_F}{2\pi^2q} \int_0^\infty dk \,
  k \, n_k \, e^{-2\left[\int_0^\ell \omega_q^2(\ell')
      d\ell' - \ell\bar\omega_q^2(\ell)\right]} \notag \\
  & \quad \times \left\{
    \frac{e_-^+ - e_+^+ + e_-^- - e_+^-}{4\ell} \right. \notag \\
  \label{eq:oflow}
  & \qquad \quad \left. + (\bar\omega_q - \omega_q) \,
    \frac{E_-^+ - E_+^+ - E_-^- + E_+^-}{2\sqrt{2\ell/\pi}} \right\}
\end{align}
with $\bar\omega_q(\ell) = (1/\ell) \int_0^\ell \omega_q(\ell')
d\ell'$, $e_\sigma^{\sigma'} = \exp[-2\ell
(\alpha_\sigma^{\sigma'})^2]$, \linebreak $E_\sigma^{\sigma'} =
\erf[\sqrt{2\ell}\alpha_\sigma^{\sigma'}]$ and
$\alpha_\sigma^{\sigma'} = q^2/(2m_F) +\sigma kq/m_F +
\sigma'\bar\omega_q$.  We use a Sommerfeld expansion for the
temperature dependence of the Fermi function $n_k$ which allows the
remaining $k$ integrals to be evaluated analytically.  The flow
equation is then integrated numerically for different $q$ values; we
use a logarithmic $q$ grid near the phase transition.

In order to determine the transition temperature for fermion pairing,
we concentrate on the BCS channel of the induced interaction,
$V_{kq}^{\text{BCS}} = V_{k,-k,q}$, for which the flow equation
simplifies to
\begin{align}
  \label{eq:Vflow}
  \frac{dV_{kq}^{\text{BCS}}}{d\ell} & =
  \left(\beta_{kq}-\alpha_{kq}\right) M_{kq} M_{k+q,-q} \\
  & = -2\omega_q M_q^2 \exp\Bigl(-\int_0^\ell
  (\alpha_{kq}^2+\beta_{kq}^2)
  d\ell'\Bigr) \notag \\
  & = -2 \omega_q M_q^2 \exp\Bigl( -2\ell(\epsilon_{k+q}-\epsilon_k)^2
  - 2\int_0^\ell \omega_q^2 d\ell' \Bigr) \notag
\end{align}
If the phonon frequency is not renormalized (mean field), the flow
equation can be integrated to give
\begin{align}
  V_{kq}^{\text{BCS}}(\ell=\infty) = -\frac{\omega_q
    M_q^2}{\omega_q^2 + (\epsilon_{k+q}-\epsilon_k)^2}
\end{align}
which differs from the Fr\"ohlich result \eqref{eq:VFr} by the $+$
sign in the denominator: this induced interaction is always attractive
and vanishes for large energy transfer (retarded).  The difference is
due to using a different fermionic quasiparticle basis.

For solving the gap equation it will be useful to express the induced
interaction between a $k$, $-k$ Cooper pair and a $k'$, $-k'$ pair in
terms of energy variables $\epsilon = \epsilon_k$, $\epsilon' =
\epsilon_{k'}$ and average over the angle:
\begin{align}
  \label{eq:Vbare}
  V(\epsilon,\epsilon') & =
  -\frac{N(E_F)^2}{2k_F^2N(\epsilon)N(\epsilon')}
  \int_{\abs{k-k'}}^{k+k'} dq \, q \, \frac{\omega_q
    M_q^2}{\omega_q^2+(\epsilon'-\epsilon)^2}
\end{align}
with $k=\sqrt{2m_F(\epsilon+\mu_F)}$ and likewise for $k'$.
Specializing further to Cooper pairs on the Fermi surface one obtains
the phonon-induced coupling strength \cite{heiselberg2000iii}
\begin{align}
  \label{eq:Vfermi}
  \lambda & = -N(E_F)V(\epsilon=\epsilon'=E_F) \\
  & = \frac{N(E_F)}{2k_F^2} \int_0^{2k_F} dq \, q \,
  \frac{M_q^2}{\omega_q} \notag \\
  & = \frac{\gamma_{BF}^2}{\gamma_{BB}(2k_F\xi)^2} \ln[1+(2k_F\xi)^2]
  \notag
\end{align}
which depends only logarithmically on $\gamma_{BB}$ (see equation
\eqref{eq:lambda} above).  Inserting this expression independent of
energy into the gap equation
\begin{align}
  \label{eq:gap}
  \Delta(\epsilon) & = -\int d\epsilon' N(\epsilon')
  V(\epsilon,\epsilon') \frac{\Delta(\epsilon')}{2E(\epsilon')} 
  \tanh\left(\frac{E(\epsilon')}{2T}\right) 
\end{align}
yields the well-known weak-coupling result
\begin{align}
  \label{eq:tcweak}
  T_c & = \frac{\gamma}{\pi}\left(\frac{2}{e}\right)^{7/3}
  \exp\{-1/\lambda\}
\end{align}
where now and in the following we have already included the correction
to the prefactor due to the polarization of the fermions
\cite{gorkov1961cts}.

However, the projection onto the Fermi surface in equation
\eqref{eq:Vfermi} \cite{heiselberg2000iii} is justified only when $c_s
\gg v_F$ and becomes insufficient as $c_s \lesssim v_F$ and
retardation effects become important.  Therefore, it is necessary to
go beyond the mean-field level and include the damping of the phonons
and its influence on the induced interaction by solving the full flow
equations \eqref{eq:oflow} and \eqref{eq:Vflow}, and finally solving
the gap equation \eqref{eq:gap} with this renormalized effective
interaction.

%%%%%%%%%%%%%%%%%%%%%%%%%%%%%%%%%%%%%%%%%%%%%%%%%%%%%%%%%%%%%%%%%%%%%%%%
\section{Results}
\label{sec:results}
%%%%%%%%%%%%%%%%%%%%%%%%%%%%%%%%%%%%%%%%%%%%%%%%%%%%%%%%%%%%%%%%%%%%%%%%
The flow equation \eqref{eq:oflow} yields the renormalized phonon
spectrum due to the excitation of fermionic particle-hole pairs.  The
Bogoliubov spectrum for a given repulsive $g_{BB}>0$ is softened upon
increasing $g_{BF}^2$ up to the point where $\omega_{q=0}$ turns
negative; this signals a local instability towards phase separation.
Typical $^{40}$K-$^{87}$Rb systems
\cite{modugno2002cdf,ospelkaus2006idd} with the K atoms prepared in
the $\lvert 9/2,-9/2\rangle$, $\lvert 9/2,-7/2\rangle$ hyperfine
states and the Rb atoms in the $\lvert 1,1\rangle$ state have fermion
densities $n_F = n_F^{\uparrow,\downarrow} \simeq 10^{12}$ cm$^{-3}$
(such that $k_F = k_F^{\uparrow,\downarrow} \approx 1.7 \cdot 10^{-4}
a_0^{-1}$), a boson density $n_B \simeq 10^{14}$ cm$^{-3}$, and a mass
ratio $r = m_F/m_B = 0.46$.  The background $s$-wave scattering
lengths are $a_{BB} = 99 a_0$, $a_{BF} = -284 a_0$, $a_{FF} = 174 a_0$
\cite{ospelkaus2006idd}.  This results in the dimensionless couplings
\begin{align}
  \gamma_{BB} & = N(E_F) g_{BB} = \frac{2r}{\pi} k_F a_{BB}
  \approx 0.005 \\
  \gamma_{BF} & = N(E_F) g_{BF} = \frac{1+r}{\pi} k_F a_{BF}
  \approx -0.02 \\
  \gamma_{FF} & = N(E_F) g_{FF} = \frac{2}{\pi} k_F a_{FF}
  \approx 0.02 \,.
\end{align}
These couplings may be tuned by Feshbach resonances towards the phase
transition at $\gamma_{BF}^2=\gamma_{BB}$ (mean field).  Moreover, for
the healing length $\xi$
\begin{align}
  k_F\xi = \sqrt{\frac{k_F^2}{4n_Bm_Bg_{BB}}}
  = \sqrt{\frac{3r}{4(n_B/n_F)\gamma_{BB}}}
  \approx 0.8
\end{align}
and the ratio of phonon and fermion velocities is
\begin{align}
  \frac{c_s}{v_F} = \frac{rm_B}{k_F} \, \sqrt{\frac{n_Bg_{BB}}{m_B}}
  = \frac{r}{2k_F\xi}
  \approx 0.3 \,.
\end{align}
In this regime, corrections to the fermion-phonon vertex are small
($\Gamma<0.1$) \cite{wang2006sct,endnote3}.

\begin{figure}
  \includegraphics[width=\linewidth,clip]{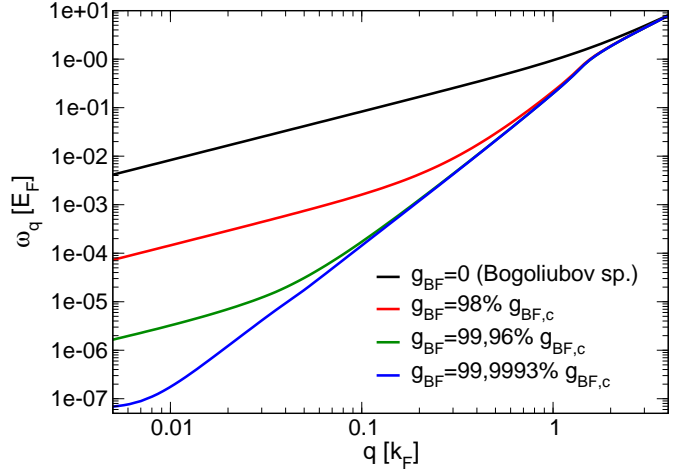}
  \caption{[color online] Renormalized phonon spectrum $\omega_q$ for
    fixed $\gamma_{BB}=0.005$ and $g_{BF}\to g_{BF,c}$ approaching the
    transition towards phase separation.  From top to bottom:
    $g_{BB}=0$ (Bogoliubov spectrum, black), $g_{BF}=0.98\, g_{BF,c}$
    (red), $g_{BF}=0.9996\, g_{BF,c}$ (green), $g_{BF}=0.999993\,
    g_{BF,c}$ (blue).}
  \label{fig:omega}
\end{figure}

The spectrum is most interesting at this critical point because here
we expect the induced fermion interaction to be largest.
Fig.~\ref{fig:omega} shows the phonon spectrum for $\gamma_{BB}$ fixed
at a realistic value $\gamma_{BB}=0.005$, while $\gamma_{BF}$ is tuned
up towards its critical value $\gamma_{BF,c}\approx 0.07$ where
$\omega_{q=0}=0$.  Already at $\gamma_{BF}=0.98 \gamma_{BF,c}$ the
phonon velocity $c_s$ is reduced by almost two orders of magnitude.
At the critical point the phonon spectrum changes dramatically from
the linear slope of the Bogoliubov spectrum to $\omega_q \sim q^3$, an
observation reminiscent of the ferromagnetic quantum critical points
with dynamical exponent $z=3$ \cite{Her76}.

Note that this spectrum of undamped oscillations belongs not to the
original phonons but to the elementary bosonic excitations of the
interacting Hamiltonian which are phonons dressed with particle-hole
excitations.  One can perform the unitary transformation backwards to
the original basis of physical fermions and phonons to obtain the
broadening of the phonon spectral function \cite{ragwitz1999fee}.

\begin{figure}
  \includegraphics[width=\linewidth,clip]{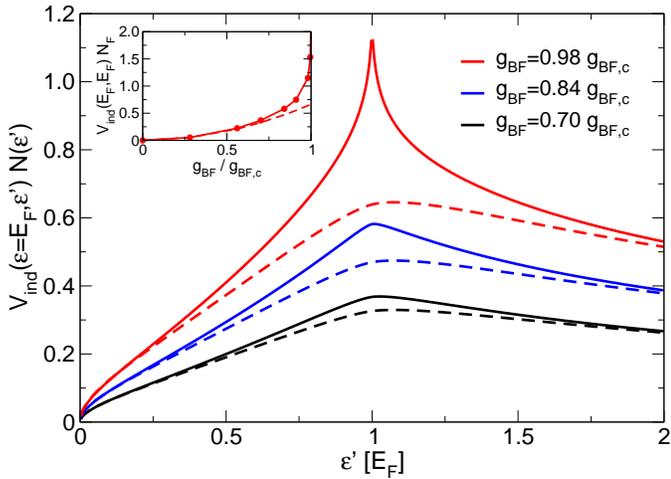}
  \caption{[color online] Induced fermion interaction in the BCS
    channel $V(\epsilon,\epsilon')$ approaching the transition towards
    phase separation.  For each pair of curves the lower (dashed) line
    is without phonon damping, while the upper (solid) line includes
    the effect of phonon damping which leads to a logarithmic
    singularity.  Parameters are $\gamma_{BB}=0.005$ and
    $g_{BF}=0.98\, g_{BF,c}$ (upper pair), $g_{BF}=0.84\, g_{BF,c}$
    (middle pair), and $g_{BF}=0.7\, g_{BF,c}$ (lower pair).  The
    inset shows the peak value of the interaction (at $\epsilon'=E_F$)
    as a function of $g_{BF}/g_{BF,c}$, again with the lower (dashed)
    line representing bare phonons and the upper (solid) line
    including renormalized phonons.}
  \label{fig:vind}
\end{figure}

The induced interaction is obtained from the flow equation
\eqref{eq:Vflow} by inserting the renormalized phonon dispersion on
the right-hand side.  Comparison with the interaction due to
unrenormalized phonons \eqref{eq:Vbare} in Fig.~\ref{fig:vind} shows
that phonon damping leads to an enhanced scattering of Cooper pairs
near the Fermi surface.  As the transition towards phase transition is
approached, this enhancement becomes more pronounced and eventually
leads to a logarithmic singularity of the induced interaction as
$\epsilon'\to E_F$, and of the peak value of the interaction as
$g_{BF}\to g_{BF,c}$ (see the inset of Fig.~\ref{fig:vind}).

\begin{figure}
  \includegraphics[width=\linewidth,clip]{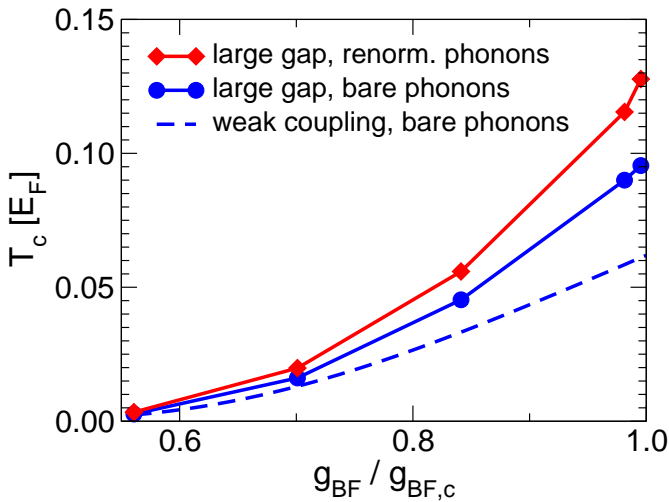}
  \caption{[color online] Transition temperature $T_c$ towards fermion
    pair superfluidity due the fermion interaction induced by phonons
    near the transition towards phase separation.  The curves are from
    top to bottom: diamonds represent the full phonon damping
    according to equation \eqref{eq:Vflow}, circles represent the
    inclusion of energies away from the Fermi level but unrenormalized
    phonons \eqref{eq:Vbare}, and the dashed line represents the
    weak-coupling result \eqref{eq:tcweak} with the interaction
    restricted to the Fermi level \eqref{eq:Vfermi}.  As before,
    $\gamma_{BB}=0.005$.}
  \label{fig:tc}
\end{figure}

We finally compare solutions of the gap equation \eqref{eq:gap} with
the different forms of the effective interaction between Cooper pairs.
By projecting all energies onto the Fermi surface one obtains the
weak-coupling result \eqref{eq:tcweak} (dashed line in
Fig.~\ref{fig:tc}), with $T_c/T_F \approx 0.05$.  Including the
dependence of $V$ on the Cooper pair energy away from the Fermi
surface but without phonon damping as in equation \eqref{eq:Vbare}
yields a somewhat higher $T_c/T_F \approx 0.08$ (circles in
Fig.~\ref{fig:tc}).  Finally, the full inclusion of phonon damping
using the flow equation \eqref{eq:Vflow} leads to a further increase
to $T_c/T_F \approx 0.1$ (diamonds).

%%%%%%%%%%%%%%%%%%%%%%%%%%%%%%%%%%%%%%%%%%%%%%%%%%%%%%%%%%%%%%%%%%%%%%%%
\section{Summary and discussion}
\label{sec:discuss}
%%%%%%%%%%%%%%%%%%%%%%%%%%%%%%%%%%%%%%%%%%%%%%%%%%%%%%%%%%%%%%%%%%%%%%%%
We have employed the Hamiltonian flow equation method to derive the
induced fermion interaction in a Bose-Fermi mixture near phase
separation beyond the mean-field approximation, and found an increase
in the resulting transition temperature towards fermion pair
superfluidity.

This involved going beyond the asymptotic solution of the phonon flow
equations \cite{lenz1996fee} to the full solution including
high-energy fluctuations; we obtained a dispersion $\omega_q \sim q^3$
for dressed phonons near the transition towards phase separation.
While it has been known that the phonon softening asymptotically leads
to a logarithmic singularity in the induced interaction
\cite{lenz1996fee} we have computed the form of the induced
interaction quantitatively for Cooper pairs also far away from the
Fermi level and found that the superfluid transition temperature near
the phase separation instability will increase substantially due to
phonon damping.

In order to identify the experimental parameters which are most
favorable for the observation of superfluidity in Bose-Fermi mixtures,
we note that the transition temperature $T_c$ grows monotonically with
the induced fermion coupling $\lambda$ in equation \eqref{eq:lambda},
both at weak and strong coupling.  For the experimental parameters
given above, $x\approx 70$.  For strong boson repulsion and/or light
fermions, $x\gamma_{BB} \gg 1$ and the coupling $\lambda\to
\lambda_{\text{MF}} = \gamma_{BF}^2 / \gamma_{BB}$ agrees with the
mean-field value \eqref{eq:lambdamf}.  Since $\lambda$ decreases
monotonically with $\gamma_{BB}$, one obtains a larger coupling
$\lambda$ by reducing $\gamma_{BB}$.  However, one eventually crosses
over into the regime where $x\gamma_{BB} \lesssim 1$ and $\lambda$ is
enhanced only logarithmically when further decreasing $\gamma_{BB}$.

At the same time, $\lambda$ is a monotonically increasing function of
$x$, so one can reach a higher critical temperature by increasing the
density of bosons relative to fermions, or the mass ratio between
bosons and fermions.  To conclude, the most direct way to observe
superfluidity appears to be to increase the strength of the Bose-Fermi
coupling $\abs{\gamma_{BF}}$, for instance by a magnetically tuned
Feshbach resonance, to the vicinity of phase separation or collapse.
It will also be interesting to study the influence of a fermion mass
difference on fermion pairing which is relevant in experiments with
$^6$Li--$^{40}$K mixtures immersed in a $^{87}$Rb condensate
\cite{taglieber2008qdt}.

\acknowledgement{This work was supported by the DFG Research Unit 801
  ``Strong correlations in multiflavor ultracold quantum gases.''}

\raggedright


\begin{thebibliography}{99}

\bibitem{group1}
%\bibitem{demarco1999ofd}
B.~DeMarco, D.S. Jin, Science \textbf{285}, 1703 (1999);
%
%\bibitem{schreck2001qbe}
F.~Schreck, L.~Khaykovich, K.L. Corwin, G.~Ferrari, T.~Bourdel, J.~Cubizolles,
  C.~Salomon, Phys.\ Rev.\ Lett.\ \textbf{87}, 080403 (2001);
%
%\bibitem{truscott2001ofp}
A.G. Truscott, K.E. Strecker, W.I. McAlexander, G.B. Partridge, R.G. Hulet,
  Science \textbf{291}, 2570 (2001);
%
%\bibitem{modugno2001bec}
G.~Modugno, G.~Ferrari, G.~Roati, R.J. Brecha, A.~Simoni, M.~Inguscio, Science
  \textbf{294}, 1320 (2001);
%
%\bibitem{roati2002fbq}
G.~Roati, F.~Riboli, G.~Modugno, M.~Inguscio, Phys.\ Rev.\ Lett. \textbf{89},
  150403 (2002)

\bibitem{taglieber2008qdt}
M.~Taglieber, A.C. Voigt, T.~Aoki, T.W. H{\"a}nsch, K.~Dieckmann, Phys.\ Rev.\
  Lett.\ \textbf{100}, 010401 (2008)

\bibitem{group2}
%\bibitem{molmer1998bca}
K.~M{\o}lmer, Phys.\ Rev.\ Lett. \textbf{80}, 1804 (1998);
%
%\bibitem{viverit2000ztp}
L.~Viverit, C.J. Pethick, H.~Smith, Phys.\ Rev.~A \textbf{61}, 053605 (2000)

\bibitem{group3}
%\bibitem{roth2002mfi}
R.~Roth, H.~Feldmeier, Phys.\ Rev.~A \textbf{65}, 021603 (2002);
%
%\bibitem{rothel2007das}
S.~R{\"o}thel, A.~Pelster, Eur.\ Phys.~J.~B \textbf{59}, 343 (2007);
%
%\bibitem{shibata2007cma}
H.~Shibata, N.~Yokoshi, S.~Kurihara, Phys.\ Rev.~A \textbf{75}, 053615 (2007)

\bibitem{modugno2002cdf}
G.~Modugno, G.~Roati, F.~Riboli, F.~Ferlaino, R.J. Brecha, M.~Inguscio, Science
  \textbf{297}, 2240 (2002)

\bibitem{ospelkaus2006idd}
C.~Ospelkaus, S.~Ospelkaus, K.~Sengstock, K.~Bongs, Phys.\ Rev.\ Lett.
  \textbf{96}, 020401 (2006)

\bibitem{endnote1}
The collapse for attractive Bose-Fermi interactions is avoided by
  taking into account the possibility of pairing between bosons and
  fermions that is neglected in mean-field theory; see F.M. Marchetti,
  C.~Mathy, M.M. Parish, D.A. Huse, Phys.\ Rev.~B \textbf{78}, 134517
  (2008)

\bibitem{group4}
%\bibitem{simoni2003mci}
A.~Simoni, F.~Ferlaino, G.~Roati, G.~Modugno, M.~Inguscio, Phys.\ Rev.\ Lett.
  \textbf{90}, 163202 (2003);
%
%\bibitem{stan2004ofr}
C.A. Stan, M.W. Zwierlein, C.H. Schunck, S.M.F. Raupach, W.~Ketterle, Phys.\
  Rev.\ Lett. \textbf{93}, 143001 (2004);
%
%\bibitem{inouye2004ohf}
S.~Inouye, J.~Goldwin, M.L. Olsen, C.~Ticknor, J.L. Bohn, D.S. Jin, Phys.\
  Rev.\ Lett. \textbf{93}, 183201 (2004);
%
%\bibitem{zhang2005elg}
J.~Zhang et al.,
%  E.G.M. van Kempen, T.~Bourdel, L.~Khaykovich, J.~Cubizolles,
%  F.~Chevy, M.~Teichmann, L.~Tarruell, S.~Kokkelman, C.~Salomon,
%  \emph{{Expansion of a lithium gas in the BEC-BCS crossover}},
  in \emph{XIX International Conference on Atomic Physics},
  edited by L.G. Marcassa, V.S. Bagnato, K.~Helmerson 
  (AIP, Melville, NY, 2005),
%  Vol. 770 of \emph{AIP Conference Proceedings}, 
  p.~228,
  \texttt{arXiv:cond-mat/0410167};
%
%\bibitem{ferlaino2006fsk}
F.~Ferlaino, C.~D{'}Errico, G.~Roati, M.~Zaccanti, M.~Inguscio, G.~Modugno,
  A.~Simoni, Phys.\ Rev.~A \textbf{73}, 40702 (2006);
%
%\bibitem{zaccanti2006cif}
M.~Zaccanti, C.~D{'}Errico, F.~Ferlaino, G.~Roati, M.~Inguscio, G.~Modugno,
  Phys.\ Rev.~A \textbf{74}, 041605 (2006);
%
%\bibitem{ospelkaus2006thi}
S.~Ospelkaus, C.~Ospelkaus, L.~Humbert, K.~Sengstock, K.~Bongs, Phys.\ Rev.\
  Lett. \textbf{97}, 120403 (2006)

\bibitem{heiselberg2000iii}
H.~Heiselberg, C.J. Pethick, H.~Smith, L.~Viverit, Phys.\ Rev.\ Lett.
  \textbf{85}, 2418 (2000)

\bibitem{group4a}
%\bibitem{bijlsma2000ped}
M.J. Bijlsma, B.A. Heringa, H.T.C. Stoof, Phys.\ Rev.~A \textbf{61}, 053601
  (2000);
%
%\bibitem{illuminati2004hta}
F.~Illuminati, A.~Albus, Phys.\ Rev.\ Lett. \textbf{93}, 090406 (2004)

\bibitem{group5}
%\bibitem{efremov2002pwc}
D.V. Efremov, L.~Viverit, Phys.\ Rev.~B \textbf{65}, 134519 (2002);
%
%\bibitem{suzuki2008pws}
K.~Suzuki, T.~Miyakawa, T.~Suzuki, Phys.\ Rev.~A \textbf{77}, 043629 (2008)

\bibitem{kalas2008ofp}
R.M. Kalas, A.V. Balatsky, D.~Mozyrsky, Arxiv preprint arXiv:0806.0419 (2008)

\bibitem{group6}
%\bibitem{buchler2003svp}
H.P. B{\"u}chler, G.~Blatter, Phys.\ Rev.\ Lett. \textbf{91}, 130404 (2003);
%
%\bibitem{buchler2004psa}
H.P. B{\"u}chler, G.~Blatter, Phys.\ Rev.~A \textbf{69}, 063603 (2004)

\bibitem{lewenstein2004abf}
M.~Lewenstein, L.~Santos, M.A. Baranov, H.~Fehrmann, Phys.\ Rev.\ Lett.
  \textbf{92}, 050401 (2004)

\bibitem{ashcroft1976ssp}
N.~Ashcroft, N.~Mermin, \emph{{Solid state physics}} (Holt, Rinehart and
  Winston, New York, 1976)

\bibitem{frohlich1952iel}
H.~Fr{\"o}hlich, Proc.\ Roy.\ Soc.\ (London)~A \textbf{215}, 291 (1952)

\bibitem{endnote2}
Note that the effective potential becomes instantaneous only in the
  limit $c_s\gg v_F$, while retardation effects are still negligible
  in the case $c_s\ll v_F$ relevant for electronic superconductivity
  because $T_c/T_F$ is exponentially small.

\bibitem{ketterle2008mpa}
W.~Ketterle, M.~Zwierlein, 
%  \emph{{Making, probing and understanding ultracold Fermi gases}},
  in \emph{Ultracold Fermi Gases, Proceedings of the
  International School of Physics ``Enrico Fermi'', Varenna},
  edited by M.~Inguscio, W.~Ketterle, C.~Salomon
  (IOS Press, Amsterdam, 2008), 
  \texttt{arXiv:0801.2500}

\bibitem{bloch2008mbp}
I.~Bloch, J.~Dalibard, W.~Zwerger, Rev.\ Mod.\ Phys.\ \textbf{80}, 885 (2008)

\bibitem{giorgini2008tuf}
S.~Giorgini, L.~Pitaevskii, S.~Stringari, Rev.\ Mod.\ Phys.\ \textbf{80}, 1215
  (2008)

\bibitem{abrikosov1963mqf}
A.A. Abrikosov, L.P. Gorkov, I.E. Dzyaloshinski, \emph{{Methods of quantum
  field theory in statistical physics}} (Prentice-Hall, Englewood
  Cliffs, NJ, 1963)

\bibitem{wang2006sct}
D.W. Wang, Phys.\ Rev.\ Lett. \textbf{96}, 140404 (2006)

\bibitem{wegner1994feh}
F.~Wegner, Ann.\ Physik (Leipzig) \textbf{3}, 77 (1994)

\bibitem{kehrein2006fea}
S.~Kehrein, \emph{{The Flow Equation Approach to Many-Particle Systems}}
  (Springer, Berlin, 2006)

\bibitem{lenz1996fee}
P.~Lenz, F.~Wegner, Nucl.\ Phys.~B \textbf{482}, 693 (1996)

\bibitem{mielke1997cct}
A.~Mielke, Europhys.\ Lett.\ \textbf{40}, 195 (1997)

\bibitem{pethick2002bec}
C.~Pethick, H.~Smith, \emph{{Bose-Einstein Condensation in Dilute Gases}}
  (Cambridge University Press, Cambridge, 2002)

\bibitem{mahan2000mpp}
G.D. Mahan, \emph{{Many-Particle Physics}} (Plenum, New York, 2000)

\bibitem{inprep}
T.~Enss, W.~Zwerger (in preparation)

\bibitem{gorkov1961cts}
L.P. Gorkov, T.K. Melik-Barkhudarov, Zh.\ Eksp.\ Teor.\ Fiz.\ \textbf{40}
  (1961)

\bibitem{endnote3} 
Vertex corrections will become important at the transition towards
  phase separation but as in \cite{kalas2008ofp} we assume that
  slightly away from the transition they are still small enough.

\bibitem{Her76}
J.~Hertz, Phys.\ Rev.~B \textbf{14}, 1165 (1976)

\bibitem{ragwitz1999fee}
M.~Ragwitz, F.~Wegner, Eur.\ Phys.~J.~B \textbf{8}, 9 (1999)

\end{thebibliography}
\end{document}